\documentclass[%
reprint,
superscriptaddress,
amsmath,amssymb,
aps,
pra
]{revtex4-2}

\usepackage[utf8]{inputenc}
\usepackage{graphicx}
\usepackage{dcolumn}
\usepackage{bm}
\usepackage{soul}
\usepackage[colorlinks]{hyperref}

\usepackage[english]{babel}
\usepackage{color}
\usepackage{mathtools}
\usepackage{cleveref}
\usepackage{verbatim}
\usepackage{breqn}

\makeatletter
\newcommand{\customlabel}[2]{%
	\protected@write \@auxout {}{\string \newlabel {#1}{{#2}{\thepage}{#2}{#1}{}} }%
	\hypertarget{#1}{}
}
\makeatother

\begin{document}
	
	\title{Quartic Kerr cavity combs: Bright and dark solitons}
	\author{Pedro Parra-Rivas}
	\affiliation{Dipartimento di Ingegneria dell’Informazione$,$ Elettronica e Telecomunicazioni$,$
		Sapienza Universit\'a di Roma$,$ via Eudossiana 18$,$ 00184 Rome$,$ Italy}
	\email{pedro.parra-rivas@uniroma1.it,carmien@upvnet.upv.es}
	\author{Sabrina Hetzel}
	\affiliation{Department of Mathematics$,$ Southern Methodist University$,$ Dallas$,$ TX 75275$,$ USA}
	
	\author{Yaroslav V. Kartashov}
	\affiliation{Institute of Spectroscopy$,$ Russian Academy of Sciences$,$ Troitsk$,$ Moscow$,$ 108840$,$ Russia}

	\author{Pedro Fern\'andez de C\'ordoba}
	\affiliation{Institut Universitari de Matem\`{a}tica Pura i Aplicada$,$ Universitat Polit\`{e}cnica de Val\`{e}ncia$,$ 46022 Val\`{e}ncia$,$ Spain}

	\author{J. Alberto Conejero}
	\affiliation{Institut Universitari de Matem\`{a}tica Pura i Aplicada$,$ Universitat Polit\`{e}cnica de Val\`{e}ncia$,$ 46022 Val\`{e}ncia$,$ Spain}

	\author{Alejandro Aceves}
	\affiliation{Institut Universitari de Matem\`{a}tica Pura i Aplicada$,$ Universitat Polit\`{e}cnica de Val\`{e}ncia$,$ 46022 Val\`{e}ncia$,$ Spain}

	\author{Carles Mili\'{a}n}
	\affiliation{Institut Universitari de Matem\`{a}tica Pura i Aplicada$,$ Universitat Polit\`{e}cnica de Val\`{e}ncia$,$ 46022 Val\`{e}ncia$,$ Spain}
	   \email{carmien@upvnet.upv.es}
	
\date{\today}

\begin{abstract}
We theoretically investigate the dynamics, bifurcation structure and stability of localized states in Kerr cavities driven at the pure fourth-order dispersion point. Both the normal and anomalous group velocity dispersion regimes are analysed, highlighting the main differences with the standard second order dispersion case. In the anomalous regime, single and multi-peak localised states exist and are stable over a much wider region of the parameter space. In the normal dispersion regime, stable narrow bright solitons exist. Some of our findings can be understood by a new scenario reported here for the spatial eigenvalues, which imposes oscillatory tails to all localised states. To be published in Optics Letters.
\end{abstract}

\maketitle



The possibility to excite robust temporal solitons in fiber loops \cite{leo_temporal_2010} and microring resonators \cite{herr_temporal_2014,xue_mode-locked_2015} had a profound impact into fundamentals and applications of frequency combs \cite{pasquazi_micro-combs:_2018,kippenberg_microresonator-based_2011}. A clear advantage of the driven systems with respect to single pass ones is the fact that ultra-short solitonic pulses exist and are robust under strong perturbations arising due to the so-called higher order effects, both of linear \cite{milianOE,parra-rivas_third-order_2014,brasch_photonic_2016} and nonlinear \cite{milian_solitons_2015,parra-rivas_influence_2021,karpov19} nature, as well as in higher dimensions \cite{milianPRL2018,IvarsPRL}. Not surprisingly, dispersion engineering was shown to be a versatile tool for control of many morphological properties of the solitonic combs, such as the enhancement of their spectral extent \cite{Okawachi_OL_14,Pfeiffer_optica_17}. In this respect, generalisations of the Schr\"{o}dinger solitons to waveguides with dominant quartic dispersion - pure quartic solitons (PQS) \cite{karlsFOD93,redNC} - are attractive because of their flatter spectra and favourable energy scaling with pulse duration, as demonstrated with mode-locked lasers \cite{redNP} and predicted also for microring resonators \cite{tahOL19}. These properties make them potentially important to increase the power of ultra-short solitons from high-$Q$ microrings. PQS's in microrings were predicted in \cite{Bao_josab_2017} and shown to be robust under Raman scattering \cite{yangOL21}.


In this letter we report on the bifurcation structure of PQSs in Kerr cavities with normal and anomalous group velocity dispersion (GVD).
 When GVD is anomalous, single and multi-peak bright localized states (LSs) [dissipative solitons], are found to coexist over the entire parameter space as well as to remain robust over a much larger area of the parameter region than in the second order dispersion system described by the Lugiato-Lefever equation (LLE) \cite{lugiato_spatial_1987}. 
 With normal GVD, stable dark and bright solitons (BSs) exist. The latter refer to the widest dark solitons and, contrarily to the LLE, are stable over a finite interval of parameters. BSs can be predicted by dynamical system theory, which ubiquitously imposes oscillatory tails to all PQSs. 

\begin{figure*}[t]
	\centering
\includegraphics[scale=1]{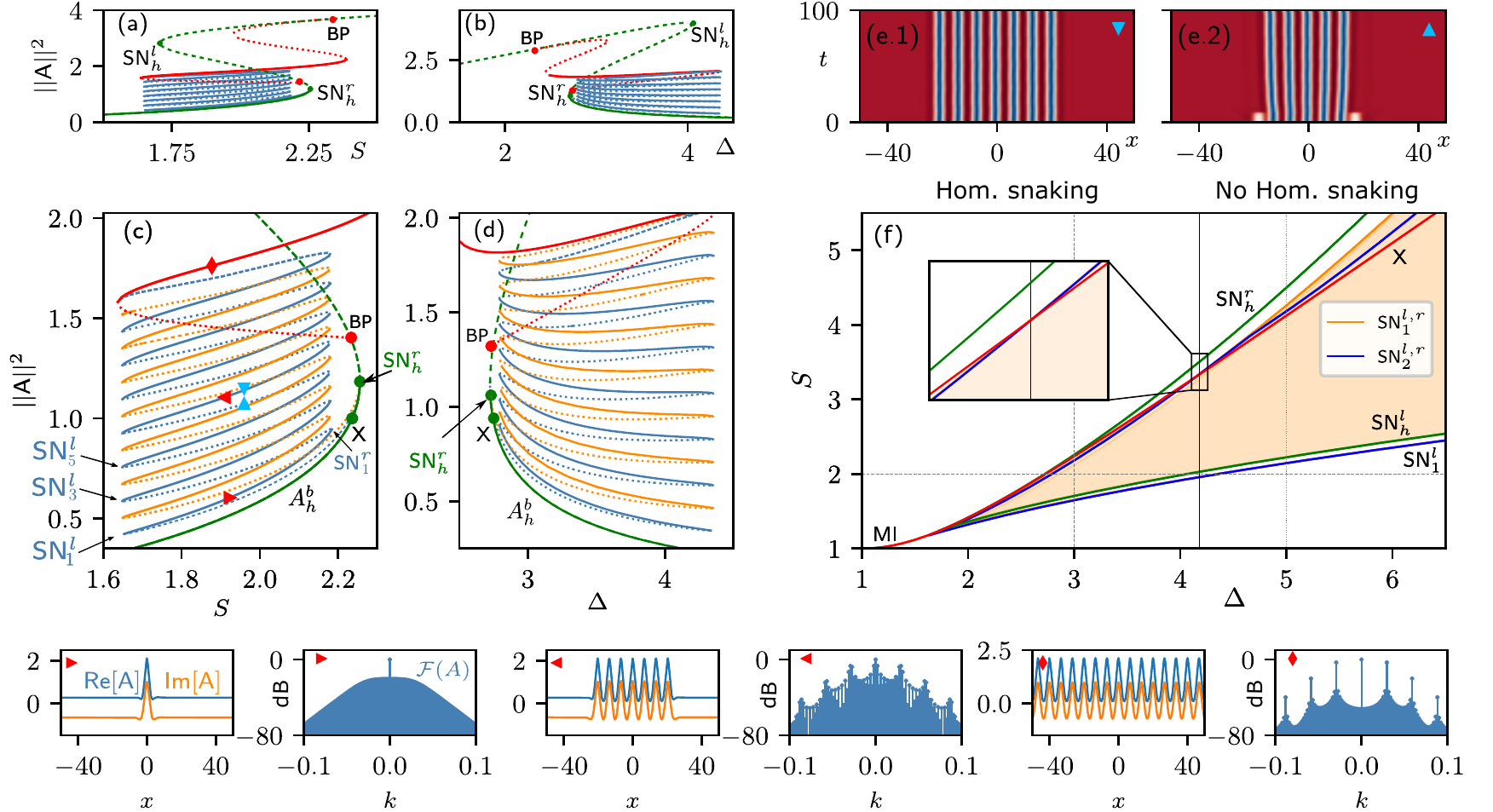}
\caption{Norm branches for the HSS (green), LSs (blue, orange), and a pattern family (red) for $\omega_4=1$ and (a,c) $\Delta=3$, (b,d) $S=2$. (c,d) zoom around the snaking curves in (a,b). Solid (dashed) lines denote stability (instability), and the red symbols correspond to the states shown in the insets. Blue symbols refer to the evolution of a stable and unstable multi-peak LS shown in (e.1) and (e.2), respectively. (f) Phase diagram in the $(\Delta,S)$-parameter space, showing the main bifurcation lines. Straight dashed gray lines at $S=2$, $\Delta=3$, and $\Delta=5$ mark, respectively, the phase space locations of the data shown in (a,c), (b,d), and Fig.\ref{fig2}(a). The inset in (f) shows the crossing (see vertical line) between the BD-analogue (red) transition and the SN$_2^r$ (blue), marked by the vertical solid line. }
	\label{fig1}
\end{figure*}
%
\begin{figure}[t]
\centering
\includegraphics[scale=1]{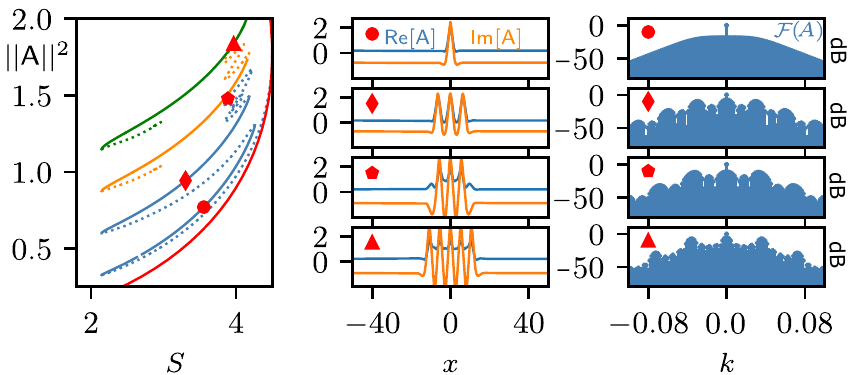}
\caption{(Left column) Disconnected standard homoclinic snaking branches for LS families with an odd number of peaks and $\Delta=5$. (Center column) Selected LS profiles and (right) associated frequency spectra.}
	\label{fig2}
\end{figure}
%
\begin{figure*}[t]
\centering
\includegraphics[scale=1]{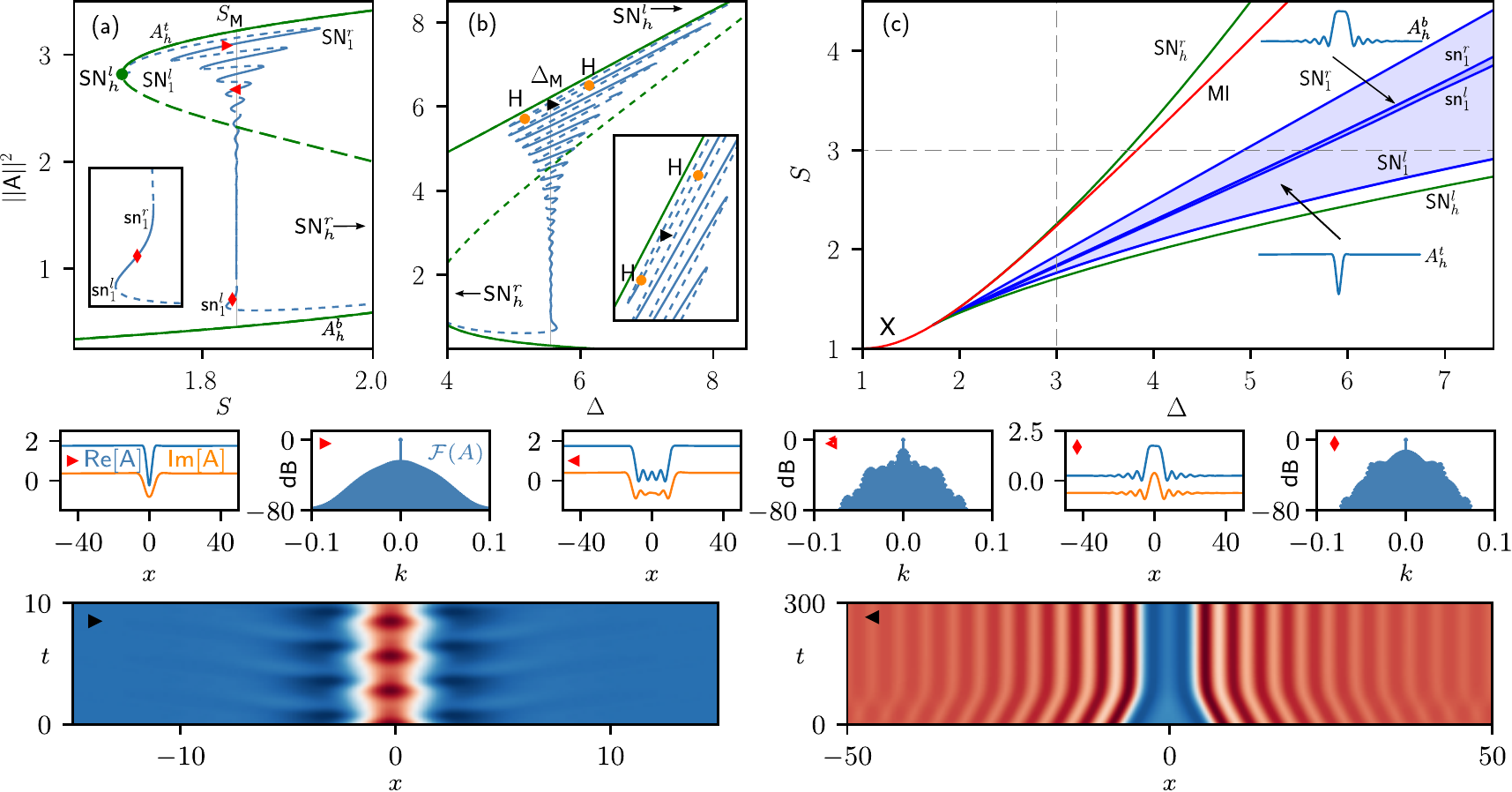}
\caption{Collapsed homoclinic snaking (blue) for $\omega_4=-1$ and (a) $\Delta=3$, (b) $S=3$. The multi-valued HSS is also shown (green). Solid (dashed) lines correspond to stable (unstable) states. Red symbols in (a) correspond to the LSs shown in the insets. H labels in (b) mark Hopf bifurcations. (c) phase diagram on the ($\Delta,S$)-plane showing the main bifurcations of the system. Bottom insets show excitation of unstable dark (left) and stable bright (right) states [see markers for reference].}
\label{fig3}         
\end{figure*}
%

Under a dominant fourth order dispersion (FOD), the time evolution of the intra-cavity field envelope, $A$, may be described by the pure quartic Lugiato-Lefever (PQLL) equation:
\begin{equation}
\partial_t A=-(1+i\Delta)A-i\omega_4\partial_x^4A+iA|A|^2+S.\label{eq1}
\end{equation}
Here, the periodic coordinate, $x$, is related to the physical length, $X$, 
by $x=X\sqrt{\gamma}/[2\pi R\sqrt{|B_4|}]$, where $R$ is the cavity radius, $\gamma$ is the normalised loss, and $B_4$ is the FOD coefficient. 

The spectral combs in this work are shown as a function of the wavenumber $k=(m-m_0)/R$, where $m_0$ is the cavity mode excited by $S$
(see, e.g., Ref.\cite{IvarsPRL} for scaling). In Eq.~(\ref{eq1}), $\omega_4$ takes de value $+1$ ($-1$) to account for parabolic anomalous (normal) GVD. We note that such seemingly idealised GVD landscapes are in fact attainable with $\omega_4=\pm1$ around the telecom band with fibers \cite{hooper,heidt} and microrings \cite{yaoOE21}. Residual second and third order dispersion are not expected to play a significant role as long as GVD sign remains constant. Equation \ref{eq1} is integrated in time with 4th order Runge-Kutta while stationary states and their stability are computed with Newton-Raphson and the corresponding Jacobian. 

We first focus on the anomalous GVD case ($\omega_4=1$). The energy, defined as the $L_2$-norm $||A||^2=l^{-1}\int_{-l/2}^{l/2}|A(x)|^2dx$ (with $l=2\pi R$), of the homogeneous steady state (HSS) $A_h$, is shown in Figs.\ref{fig1}(a),\ref{fig1}(b) using solid (dashed) lines for stable (unstable) states (green curves). For these parameters the HSS undergoes a pair of saddle-node bifurcations that we label SN$_h^{l,r}$. For norms above SN$_h^r$, many unstable families of Turing (periodic) patterns bifurcate sub-critically from the branching point (BP) on the HSS. Examples of such branches are plotted in Figs.~\ref{fig1}(a), \ref{fig1}(b) (red curve) and in the close-up views Figs.~\ref{fig1}(c), \ref{fig1}(d). An example of one such pattern is shown in Figs.~\ref{fig1}{\color{red}$\blacklozenge$}, in spatial and Fourier domains. Because patterns and HSS, both stable, coexist over a finite interval in pump amplitude, $S$, [Figs.~\ref{fig1}(a),(c)] or detuning, $\Delta$ [Figs.~\ref{fig1}(b),(d)], LSs consisting of a slug of the pattern embedded in a HSS [i.e., partly pattern, partly flat state], may form due to the locking of fronts \cite{woods_heteroclinic_1999}. The norm of LSs is shown in Figs.~\ref{fig1}(c),(d) by blue (orange) curves corresponding to LSs with an odd (even) number of peaks. Examples of LSs are represented in Figs.~\ref{fig1}{\color{red}$\blacktriangleright$} and \ref{fig1}{\color{red}$\blacktriangleleft$}. LS branches at higher $||A||^2$ values correspond to spatially wider states because two pattern peaks nucleate symmetrically on each side of the LSs when branches fold back towards smaller pump [larger detuning] across the saddle-nodes depicted in Fig.\ref{fig1}(c)[(d)]. This back-and-forth snaking of the LS branches corresponds to the standard homoclinic snaking \cite{woods_heteroclinic_1999}. The evolution of a stable [unstable] multi-peak state is shown in Fig.~\ref{fig1}(e.1) [(e.2)], respectively. 

Figure~\ref{fig1}(f) depicts relevant bifurcations in $(\Delta,S)$-parameter space and shows the existence domain of LSs (shaded area). The lower HSS state, $A_h^b$, becomes unstable for $S$ values above the red line (for which $I_h\equiv||A_h||^2=1$), which corresponds to modulational instability (MI) when $\Delta<2$.

In the LL system, the homoclinic snaking shown in Fig.\ref{fig1} 
ceases to exist for $\Delta\geq2$ because
(for $\Delta=2$)
the right [left] most saddle-nodes of the LSs, SN$_i^r$, shown in Fig.~\ref{fig1}(c)[(d)], collide with a homoclinic bifurcation occurring at a Belyakov-Devaney transition (marked as X) \cite{champneys_homoclinic_1998}
. As a result, single-peak BSs connect with spatially equispaced multi-peak LSs, forming a {\it foliated snaking}
\cite{parra-rivas_bifurcation_2018,qi_dissipative_2019}. 
On the contrary, in the PQLL system, the SN$_i^r$ collides with the point $X$ at a larger detuning, $\Delta\approx4.15$ [c.f. Fig.\ref{fig1}(f)], thus considerably broadening the region where standard homoclinic snaking occurs. Above $\Delta\approx4.15$, the homoclinic snaking  breaks up, 
yielding bifurcation curves of the type shown in Fig.~\ref{fig2} for $\Delta=5$, which do not seem to \textit{connect} all LSs with odd (nor even) number of peaks. Interestingly, in this high detuning region, ($\Delta>4.15)$, LSs still exist [c.f. Fig.\ref{fig2}]. 
The much larger region of the phase space in which LSs exist in the PQLL system provides a unique scenario to observe these solitary waves.

We now turn to the normal GVD case ($\omega_4=-1$). Because the HSS's and corresponding stability thresholds are identical to those of the LLE, the system presents two uniform states that coexist and are stable: the bottom ($A_h^b$) and top ($A_h^t$) branches. In this situation, plane fronts, domain walls or switching waves (SW) may form, interact and bind together leading to the formation of LSs. Our findings are summarised in Fig.~\ref{fig3}. Similarly to the LL system,
dark solitons bifurcate from SN$_h^l$ and undergo collapsed homoclinic snaking \cite{knobloch_homoclinic_2005}: an oscillatory bifurcation curve that damps to the uniform Maxwell point ($S_M$ or $\Delta_M$) of the system as $||A||^2$ decreases [c.f. Figs.~\ref{fig3}(a) and \ref{fig3}(b)]. Selected LSs are shown in Figs.~\ref{fig3}{\color{red}$\blacktriangleright$} and \ref{fig3}{\color{red}$\blacktriangleleft$}. LSs existing between two SN bifurcations often present Hopf instability [c.f. Fig.\ref{fig3}(b)] leading to the appearance of breathers [c.f. Fig.~\ref{fig3}$\blacktriangleright$]. 

The collapsed snaking in Figs.~\ref{fig3}(a) and \ref{fig3}(b) is, thus far, qualitatively identical to that of the LL system \cite{lobanov_frequency_2015,parra-rivas_dark_2016,parra-rivas_coexistence_2017}. However, the most evident and interesting distinctive feature here is the presence of a stable branch at the bottom of the diagram, shown in close-up view in Fig.~\ref{fig3}(a), limited by sn$_1^{l,r}$, which does not exist in the LLE. The LSs along this branch are stable BSs, similar to those found under other higher order effects \cite{parra-rivas_coexistence_2017,li_experimental_2020,parra-rivas_influence_2021,mbe17} and reminiscent of the narrowest platicons found in \cite{lobanov_frequency_2015} (c.f. Fig.~\ref{fig3}{\color{red}$\blacklozenge$}). Figure~\ref{fig3}(c) shows reference bifurcations on the $(\Delta,S)$ plane. HSS bistability exists between SN$_h^l$ and SN$_h^r$ (see green curves), $A_h^b$ suffers MI in the region between the red curve and SN$_h^r$. The narrowest and stable dark LSs exist within the light shaded region (limited by SN$_1^{r,l}$) and BSs are found within the narrowest shaded region (between sn$_1^l$ and sn$_1^r$).


The origin of these narrow BSs can be formally understood by means of dynamical system's theory \cite{wiggins_introduction_2003}. Briefly, defining $U=\mathrm{Re}[A(x)]$, $V=\mathrm{Im}[A(x)]$, and $y(x)=[U,U',U'',U''',V,V',V'',V''']^T$, it is possible to cast Eq.~(\ref{eq1}) in the form $y'(x)=f(y(x);\Delta,S,\omega_4)$, which linearised around a HSS $y_0$, reads $y'(x)=\mathcal{J}(y_0)y(x)+\mathcal{O}(y^2(x))$. The eigenvalues, $\lambda$, of the Jacobian matrix $\mathcal{J}(y_0)$ determine whether the tails of a stationary state (SWs or otherwise) are oscillatory ($\mathrm{Im}[\lambda]\neq0$) or monotonic ($\mathrm{Im}[\lambda]=0$) and solve the bi-quartic equation
\begin{equation}
    \lambda^8-(4I_h-2\Delta)\omega_4\lambda^4+3I_h^2-4\Delta I_h+\Delta^2+1=0.
    \label{eq2}
\end{equation}
This equation transforms into that corresponding to the LLE by substituting $\lambda\rightarrow\pm\sqrt{i\lambda_{LL}}$ and $\omega_4\rightarrow\omega_2$ ($\omega_2$ is the GVD coefficient). With this simple connection between the LL and the PQLL systems (and taking into account that the HSS are exactly the same in both cases), it is straightforward to relate the eigenvalue distributions of both in a general way.

\begin{figure}[t]
\centering
\includegraphics[scale=0.34]{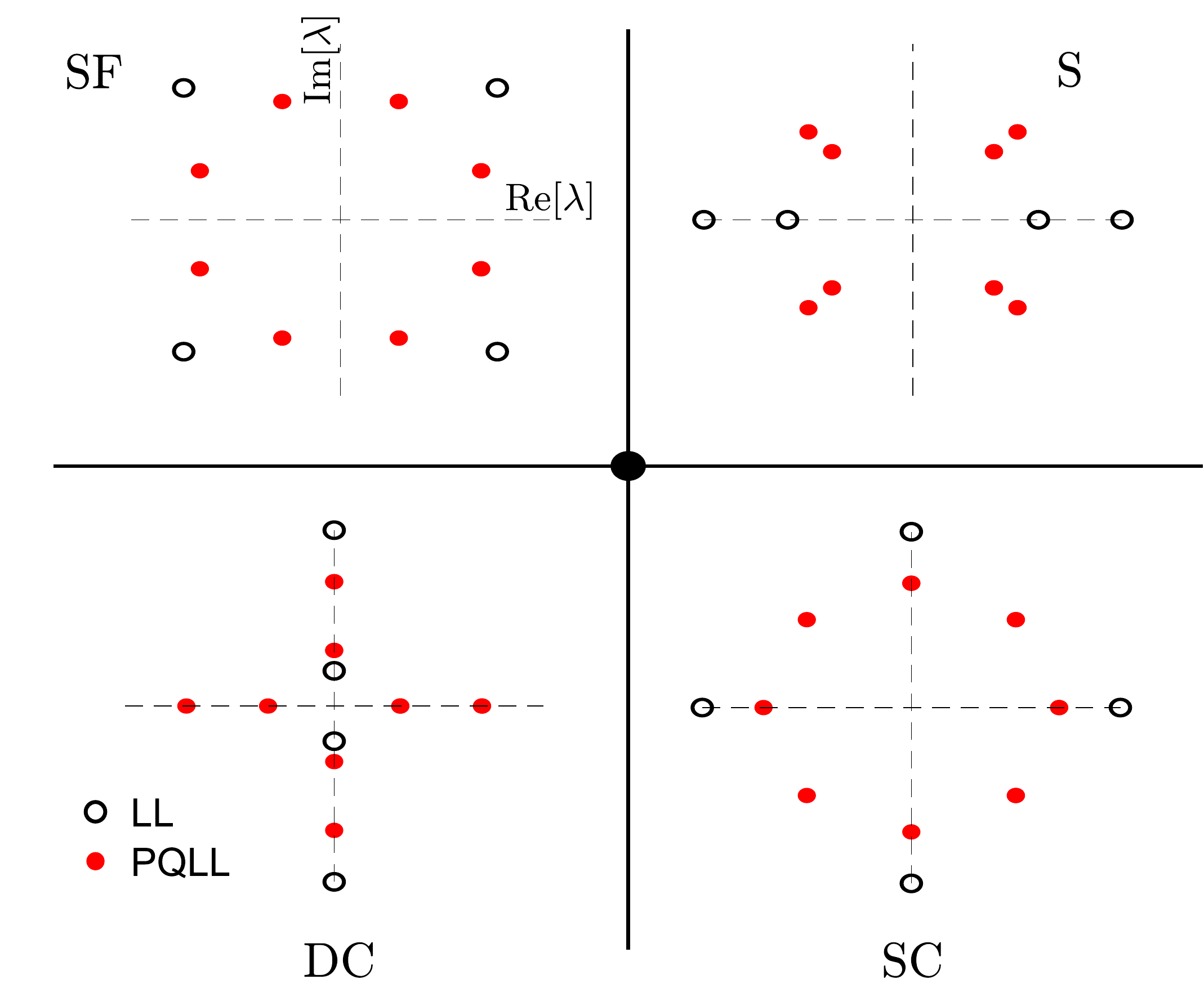}
\caption{
General picture of the spatial eigenvalue distributions for the standard LL (black hollow circles) and the PQLL [Eq.~(\ref{eq1})] (red filled dots) equations.
}
\label{fig4}         
\end{figure}
%
\begin{figure}[t]
\centering
\includegraphics[scale=0.85]{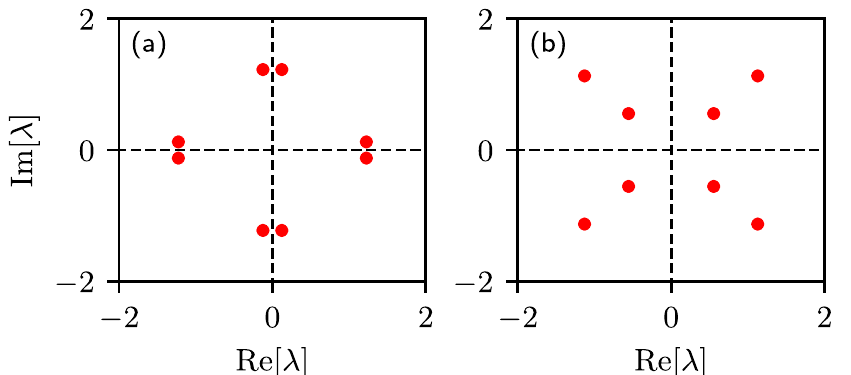}
\caption{The spatial eigenvalues associated to the bright soliton in Fig.\ref{fig3}{\color{red}$\blacklozenge$} are shown for (a) the bottom state $A_h^b$ with $(\omega_4,\Delta,I_h)=(-1,3,0.44)$ and (b) the top state $A_h^t$ with $(\omega_4,\Delta,I_h)=(-1,3,3.22)$.}
\label{fig5}         
\end{figure}
In the LLE, four different types of eigenvalue distributions corresponding to saddle (S), saddle-focus (SF), double center (DC), and saddle-center (SC) fixed points, are known \cite{parra-rivas_bifurcation_2018-1} (hollow circles in Fig.~\ref{fig4}). The relation $\lambda\rightarrow\pm\sqrt{i\lambda_{LL}}$ allows us to visualize the eigenvalue distributions of the PQLL (see red filled dots in Fig.~\ref{fig4}). 

SWs connect the bottom and top HSSs ($A_h^{b,t}$), which in the LLE are always an SF and an S, respectively \cite{parra-rivas_dark_2016}. Hence, oscillations only exist on the lower state of the SW. In that case, short dark LSs form easily while short bright LS (or, equivalently, extremely long dark LS) struggle to form. On the contrary, in the PQLL, the saddle (S) fixed point becomes a quatri-focus (top-right panel in Fig.~\ref{fig4}) with nonzero $\mathrm{Im}\{\lambda\}$. This induces oscillations on the top HSS, $A_h^t$, and the BSs appear due to the locking of plane fronts around such states.
In particular, the set of spatial eigenvalues shown in Figs.~\ref{fig5}(a) and \ref{fig5}(b) correspond to the lower and upper states of the SWs that bind together to form the BS in Fig.~\ref{fig3}{\color{red}$\blacklozenge$}. The process of interaction of two SWs around $A_h^t$ and the locking into a BS is evident from Fig.~\ref{fig3}$\blacktriangleleft$.


To conclude, we presented the bifurcation structure, existence and stability of LSs in the PQLL system with normal and anomalous GVD. In the anomalous case, LSs exist as stable states over a much larger region than in the LL system, 
In the normal GVD regime, stable BSs exist due to SW locking over their tops. Some of the qualitative differences found here with respect to the LLE can be formally understood by dynamical system's theory.\\
{\bf Funding.} Division of Mathematical Sciences and National Science Foundation(1909559); Division of Mathematical Sciences (1840260); Research Executive Agency (101023717).\\
{\bf Acknowledgment.} PPR
acknowledges the European Union’s Marie Sklodowska-Curie grant 101023717. SH and AA  have been supported by the NSF RTG grant DMS-1840260  and NSF/DMS-1909559 (AA). They also benefited from discussions with Dr. Ross Parker. YVK and CM gratefully acknowledge discussions with Prof. L. Torner.\\ 
{\bf Disclosures.}
The authors declare no conflicts of interest.\\
{\bf Data availability.} Data underlying the results in this paper may be obtained from the authors upon reasonable request.

\bibliography{LSs_SHG}

\begin{thebibliography}{37}%
\makeatletter
\providecommand \@ifxundefined [1]{%
 \@ifx{#1\undefined}
}%
\providecommand \@ifnum [1]{%
 \ifnum #1\expandafter \@firstoftwo
 \else \expandafter \@secondoftwo
 \fi
}%
\providecommand \@ifx [1]{%
 \ifx #1\expandafter \@firstoftwo
 \else \expandafter \@secondoftwo
 \fi
}%
\providecommand \natexlab [1]{#1}%
\providecommand \enquote  [1]{``#1''}%
\providecommand \bibnamefont  [1]{#1}%
\providecommand \bibfnamefont [1]{#1}%
\providecommand \citenamefont [1]{#1}%
\providecommand \href@noop [0]{\@secondoftwo}%
\providecommand \href [0]{\begingroup \@sanitize@url \@href}%
\providecommand \@href[1]{\@@startlink{#1}\@@href}%
\providecommand \@@href[1]{\endgroup#1\@@endlink}%
\providecommand \@sanitize@url [0]{\catcode `\\12\catcode `\$12\catcode
  `\&12\catcode `\#12\catcode `\^12\catcode `\_12\catcode `\%12\relax}%
\providecommand \@@startlink[1]{}%
\providecommand \@@endlink[0]{}%
\providecommand \url  [0]{\begingroup\@sanitize@url \@url }%
\providecommand \@url [1]{\endgroup\@href {#1}{\urlprefix }}%
\providecommand \urlprefix  [0]{URL }%
\providecommand \Eprint [0]{\href }%
\providecommand \doibase [0]{https://doi.org/}%
\providecommand \selectlanguage [0]{\@gobble}%
\providecommand \bibinfo  [0]{\@secondoftwo}%
\providecommand \bibfield  [0]{\@secondoftwo}%
\providecommand \translation [1]{[#1]}%
\providecommand \BibitemOpen [0]{}%
\providecommand \bibitemStop [0]{}%
\providecommand \bibitemNoStop [0]{.\EOS\space}%
\providecommand \EOS [0]{\spacefactor3000\relax}%
\providecommand \BibitemShut  [1]{\csname bibitem#1\endcsname}%
\let\auto@bib@innerbib\@empty
\bibitem [{\citenamefont {Leo}\ \emph {et~al.}(2010)\citenamefont {Leo},
  \citenamefont {Coen}, \citenamefont {Kockaert}, \citenamefont {Gorza},
  \citenamefont {Emplit},\ and\ \citenamefont
  {Haelterman}}]{leo_temporal_2010}%
  \BibitemOpen
  \bibfield  {author} {\bibinfo {author} {\bibfnamefont {F.}~\bibnamefont
  {Leo}}, \bibinfo {author} {\bibfnamefont {S.}~\bibnamefont {Coen}}, \bibinfo
  {author} {\bibfnamefont {P.}~\bibnamefont {Kockaert}}, \bibinfo {author}
  {\bibfnamefont {S.-P.}\ \bibnamefont {Gorza}}, \bibinfo {author}
  {\bibfnamefont {P.}~\bibnamefont {Emplit}},\ and\ \bibinfo {author}
  {\bibfnamefont {M.}~\bibnamefont {Haelterman}},\ }\bibfield  {title}
  {\bibinfo {title} {Temporal cavity solitons in one-dimensional {Kerr} media
  as bits in an all-optical buffer},\ }\href
  {https://doi.org/10.1038/nphoton.2010.120} {\bibfield  {journal} {\bibinfo
  {journal} {Nature Photonics}\ }\textbf {\bibinfo {volume} {4}},\ \bibinfo
  {pages} {471} (\bibinfo {year} {2010})}\BibitemShut {NoStop}%
\bibitem [{\citenamefont {Herr}\ \emph {et~al.}(2014)\citenamefont {Herr},
  \citenamefont {Brasch}, \citenamefont {Jost}, \citenamefont {Wang},
  \citenamefont {Kondratiev}, \citenamefont {Gorodetsky},\ and\ \citenamefont
  {Kippenberg}}]{herr_temporal_2014}%
  \BibitemOpen
  \bibfield  {author} {\bibinfo {author} {\bibfnamefont {T.}~\bibnamefont
  {Herr}}, \bibinfo {author} {\bibfnamefont {V.}~\bibnamefont {Brasch}},
  \bibinfo {author} {\bibfnamefont {J.~D.}\ \bibnamefont {Jost}}, \bibinfo
  {author} {\bibfnamefont {C.~Y.}\ \bibnamefont {Wang}}, \bibinfo {author}
  {\bibfnamefont {N.~M.}\ \bibnamefont {Kondratiev}}, \bibinfo {author}
  {\bibfnamefont {M.~L.}\ \bibnamefont {Gorodetsky}},\ and\ \bibinfo {author}
  {\bibfnamefont {T.~J.}\ \bibnamefont {Kippenberg}},\ }\bibfield  {title}
  {\bibinfo {title} {Temporal solitons in optical microresonators},\ }\href
  {https://doi.org/10.1038/nphoton.2013.343} {\bibfield  {journal} {\bibinfo
  {journal} {Nature Photonics}\ }\textbf {\bibinfo {volume} {8}},\ \bibinfo
  {pages} {145} (\bibinfo {year} {2014})}\BibitemShut {NoStop}%
\bibitem [{\citenamefont {Xue}\ \emph {et~al.}(2015)\citenamefont {Xue},
  \citenamefont {Xuan}, \citenamefont {Liu}, \citenamefont {Wang},
  \citenamefont {Chen}, \citenamefont {Wang}, \citenamefont {Leaird},
  \citenamefont {Qi},\ and\ \citenamefont {Weiner}}]{xue_mode-locked_2015}%
  \BibitemOpen
  \bibfield  {author} {\bibinfo {author} {\bibfnamefont {X.}~\bibnamefont
  {Xue}}, \bibinfo {author} {\bibfnamefont {Y.}~\bibnamefont {Xuan}}, \bibinfo
  {author} {\bibfnamefont {Y.}~\bibnamefont {Liu}}, \bibinfo {author}
  {\bibfnamefont {P.-H.}\ \bibnamefont {Wang}}, \bibinfo {author}
  {\bibfnamefont {S.}~\bibnamefont {Chen}}, \bibinfo {author} {\bibfnamefont
  {J.}~\bibnamefont {Wang}}, \bibinfo {author} {\bibfnamefont {D.~E.}\
  \bibnamefont {Leaird}}, \bibinfo {author} {\bibfnamefont {M.}~\bibnamefont
  {Qi}},\ and\ \bibinfo {author} {\bibfnamefont {A.~M.}\ \bibnamefont
  {Weiner}},\ }\bibfield  {title} {\bibinfo {title} {Mode-locked dark pulse
  {Kerr} combs in normal-dispersion microresonators},\ }\href
  {https://doi.org/10.1038/nphoton.2015.137} {\bibfield  {journal} {\bibinfo
  {journal} {Nature Photonics}\ }\textbf {\bibinfo {volume} {9}},\ \bibinfo
  {pages} {594} (\bibinfo {year} {2015})}\BibitemShut {NoStop}%
\bibitem [{\citenamefont {Pasquazi}\ \emph {et~al.}(2018)\citenamefont
  {Pasquazi}, \citenamefont {Peccianti}, \citenamefont {Razzari}, \citenamefont
  {Moss}, \citenamefont {Coen}, \citenamefont {Erkintalo}, \citenamefont
  {Chembo}, \citenamefont {Hansson}, \citenamefont {Wabnitz}, \citenamefont
  {Del’Haye}, \citenamefont {Xue}, \citenamefont {Weiner},\ and\
  \citenamefont {Morandotti}}]{pasquazi_micro-combs:_2018}%
  \BibitemOpen
  \bibfield  {author} {\bibinfo {author} {\bibfnamefont {A.}~\bibnamefont
  {Pasquazi}}, \bibinfo {author} {\bibfnamefont {M.}~\bibnamefont {Peccianti}},
  \bibinfo {author} {\bibfnamefont {L.}~\bibnamefont {Razzari}}, \bibinfo
  {author} {\bibfnamefont {D.~J.}\ \bibnamefont {Moss}}, \bibinfo {author}
  {\bibfnamefont {S.}~\bibnamefont {Coen}}, \bibinfo {author} {\bibfnamefont
  {M.}~\bibnamefont {Erkintalo}}, \bibinfo {author} {\bibfnamefont {Y.~K.}\
  \bibnamefont {Chembo}}, \bibinfo {author} {\bibfnamefont {T.}~\bibnamefont
  {Hansson}}, \bibinfo {author} {\bibfnamefont {S.}~\bibnamefont {Wabnitz}},
  \bibinfo {author} {\bibfnamefont {P.}~\bibnamefont {Del’Haye}}, \bibinfo
  {author} {\bibfnamefont {X.}~\bibnamefont {Xue}}, \bibinfo {author}
  {\bibfnamefont {A.~M.}\ \bibnamefont {Weiner}},\ and\ \bibinfo {author}
  {\bibfnamefont {R.}~\bibnamefont {Morandotti}},\ }\bibfield  {title}
  {\bibinfo {title} {Micro-combs: {A} novel generation of optical sources},\
  }\href {https://doi.org/10.1016/j.physrep.2017.08.004} {\bibfield  {journal}
  {\bibinfo  {journal} {Phys. Rep.}\ }\bibinfo {series} {Micro-combs: {A} novel
  generation of optical sources},\ \textbf {\bibinfo {volume} {729}},\ \bibinfo
  {pages} {1} (\bibinfo {year} {2018})}\BibitemShut {NoStop}%
\bibitem [{\citenamefont {Kippenberg}\ \emph {et~al.}(2011)\citenamefont
  {Kippenberg}, \citenamefont {Holzwarth},\ and\ \citenamefont
  {Diddams}}]{kippenberg_microresonator-based_2011}%
  \BibitemOpen
  \bibfield  {author} {\bibinfo {author} {\bibfnamefont {T.~J.}\ \bibnamefont
  {Kippenberg}}, \bibinfo {author} {\bibfnamefont {R.}~\bibnamefont
  {Holzwarth}},\ and\ \bibinfo {author} {\bibfnamefont {S.~A.}\ \bibnamefont
  {Diddams}},\ }\bibfield  {title} {\bibinfo {title} {Microresonator-{Based}
  {Optical} {Frequency} {Combs}},\ }\href
  {https://doi.org/10.1126/science.1193968} {\bibfield  {journal} {\bibinfo
  {journal} {Science}\ }\textbf {\bibinfo {volume} {332}},\ \bibinfo {pages}
  {555} (\bibinfo {year} {2011})}\BibitemShut {NoStop}%
\bibitem [{\citenamefont {Milián}\ and\ \citenamefont
  {Skryabin}(2014)}]{milianOE}%
  \BibitemOpen
  \bibfield  {author} {\bibinfo {author} {\bibfnamefont {C.}~\bibnamefont
  {Milián}}\ and\ \bibinfo {author} {\bibfnamefont {D.~V.}\ \bibnamefont
  {Skryabin}},\ }\bibfield  {title} {\bibinfo {title} {Soliton families and
  resonant radiation in a micro-ring resonator near zero group-velocity
  dispersion},\ }\href {https://doi.org/10.1364/OE.22.003732} {\bibfield
  {journal} {\bibinfo  {journal} {Optics Express}\ }\textbf {\bibinfo {volume}
  {22}},\ \bibinfo {pages} {3732} (\bibinfo {year} {2014})}\BibitemShut
  {NoStop}%
\bibitem [{\citenamefont {Parra-Rivas}\ \emph {et~al.}(2014)\citenamefont
  {Parra-Rivas}, \citenamefont {Gomila}, \citenamefont {Leo}, \citenamefont
  {Coen},\ and\ \citenamefont {Gelens}}]{parra-rivas_third-order_2014}%
  \BibitemOpen
  \bibfield  {author} {\bibinfo {author} {\bibfnamefont {P.}~\bibnamefont
  {Parra-Rivas}}, \bibinfo {author} {\bibfnamefont {D.}~\bibnamefont {Gomila}},
  \bibinfo {author} {\bibfnamefont {F.}~\bibnamefont {Leo}}, \bibinfo {author}
  {\bibfnamefont {S.}~\bibnamefont {Coen}},\ and\ \bibinfo {author}
  {\bibfnamefont {L.}~\bibnamefont {Gelens}},\ }\bibfield  {title} {\bibinfo
  {title} {Third-order chromatic dispersion stabilizes {Kerr} frequency
  combs},\ }\href {https://doi.org/10.1364/OL.39.002971} {\bibfield  {journal}
  {\bibinfo  {journal} {Optics Letters}\ }\textbf {\bibinfo {volume} {39}},\
  \bibinfo {pages} {2971} (\bibinfo {year} {2014})}\BibitemShut {NoStop}%
\bibitem [{\citenamefont {Brasch}\ \emph {et~al.}(2016)\citenamefont {Brasch},
  \citenamefont {Geiselmann}, \citenamefont {Herr}, \citenamefont {Lihachev},
  \citenamefont {Pfeiffer}, \citenamefont {Gorodetsky},\ and\ \citenamefont
  {Kippenberg}}]{brasch_photonic_2016}%
  \BibitemOpen
  \bibfield  {author} {\bibinfo {author} {\bibfnamefont {V.}~\bibnamefont
  {Brasch}}, \bibinfo {author} {\bibfnamefont {M.}~\bibnamefont {Geiselmann}},
  \bibinfo {author} {\bibfnamefont {T.}~\bibnamefont {Herr}}, \bibinfo {author}
  {\bibfnamefont {G.}~\bibnamefont {Lihachev}}, \bibinfo {author}
  {\bibfnamefont {M.~H.~P.}\ \bibnamefont {Pfeiffer}}, \bibinfo {author}
  {\bibfnamefont {M.~L.}\ \bibnamefont {Gorodetsky}},\ and\ \bibinfo {author}
  {\bibfnamefont {T.~J.}\ \bibnamefont {Kippenberg}},\ }\bibfield  {title}
  {\bibinfo {title} {Photonic chip–based optical frequency comb using soliton
  {Cherenkov} radiation},\ }\href {https://doi.org/10.1126/science.aad4811}
  {\bibfield  {journal} {\bibinfo  {journal} {Science}\ }\textbf {\bibinfo
  {volume} {351}},\ \bibinfo {pages} {357} (\bibinfo {year}
  {2016})}\BibitemShut {NoStop}%
\bibitem [{\citenamefont {Milián}\ \emph {et~al.}(2015)\citenamefont
  {Milián}, \citenamefont {Gorbach}, \citenamefont {Taki}, \citenamefont
  {Yulin},\ and\ \citenamefont {Skryabin}}]{milian_solitons_2015}%
  \BibitemOpen
  \bibfield  {author} {\bibinfo {author} {\bibfnamefont {C.}~\bibnamefont
  {Milián}}, \bibinfo {author} {\bibfnamefont {A.~V.}\ \bibnamefont
  {Gorbach}}, \bibinfo {author} {\bibfnamefont {M.}~\bibnamefont {Taki}},
  \bibinfo {author} {\bibfnamefont {A.~V.}\ \bibnamefont {Yulin}},\ and\
  \bibinfo {author} {\bibfnamefont {D.~V.}\ \bibnamefont {Skryabin}},\
  }\bibfield  {title} {\bibinfo {title} {Solitons and frequency combs in silica
  microring resonators: {Interplay} of the {Raman} and higher-order dispersion
  effects},\ }\href {https://doi.org/10.1103/PhysRevA.92.033851} {\bibfield
  {journal} {\bibinfo  {journal} {Physical Review A}\ }\textbf {\bibinfo
  {volume} {92}},\ \bibinfo {pages} {033851} (\bibinfo {year}
  {2015})}\BibitemShut {NoStop}%
\bibitem [{\citenamefont {Parra-Rivas}\ \emph {et~al.}(2021)\citenamefont
  {Parra-Rivas}, \citenamefont {Coulibaly}, \citenamefont {Clerc},\ and\
  \citenamefont {Tlidi}}]{parra-rivas_influence_2021}%
  \BibitemOpen
  \bibfield  {author} {\bibinfo {author} {\bibfnamefont {P.}~\bibnamefont
  {Parra-Rivas}}, \bibinfo {author} {\bibfnamefont {S.}~\bibnamefont
  {Coulibaly}}, \bibinfo {author} {\bibfnamefont {M.~G.}\ \bibnamefont
  {Clerc}},\ and\ \bibinfo {author} {\bibfnamefont {M.}~\bibnamefont {Tlidi}},\
  }\bibfield  {title} {\bibinfo {title} {Influence of stimulated {Raman}
  scattering on {Kerr} domain walls and localized structures},\ }\href
  {https://doi.org/10.1103/PhysRevA.103.013507} {\bibfield  {journal} {\bibinfo
   {journal} {Physical Review A}\ }\textbf {\bibinfo {volume} {103}},\ \bibinfo
  {pages} {013507} (\bibinfo {year} {2021})}\BibitemShut {NoStop}%
\bibitem [{\citenamefont {Karpov}\ \emph {et~al.}(2019)\citenamefont {Karpov},
  \citenamefont {Pfeiffer}, \citenamefont {Guo}, \citenamefont {Weng},
  \citenamefont {Liu},\ and\ \citenamefont {Kippenberg}}]{karpov19}%
  \BibitemOpen
  \bibfield  {author} {\bibinfo {author} {\bibfnamefont {M.}~\bibnamefont
  {Karpov}}, \bibinfo {author} {\bibfnamefont {M.~H.~P.}\ \bibnamefont
  {Pfeiffer}}, \bibinfo {author} {\bibfnamefont {H.}~\bibnamefont {Guo}},
  \bibinfo {author} {\bibfnamefont {W.}~\bibnamefont {Weng}}, \bibinfo {author}
  {\bibfnamefont {J.}~\bibnamefont {Liu}},\ and\ \bibinfo {author}
  {\bibfnamefont {T.~J.}\ \bibnamefont {Kippenberg}},\ }\bibfield  {title}
  {\bibinfo {title} {Dynamics of soliton crystals in optical microresonators},\
  }\href {https://doi.org/10.1038/s41567-019-0635-0} {\bibfield  {journal}
  {\bibinfo  {journal} {Nature Physics}\ }\textbf {\bibinfo {volume} {15}},\
  \bibinfo {pages} {1071} (\bibinfo {year} {2019})}\BibitemShut {NoStop}%
\bibitem [{\citenamefont {Mili{\'a}n}\ \emph {et~al.}(2018)\citenamefont
  {Mili{\'a}n}, \citenamefont {Kartashov}, \citenamefont {Skryabin},\ and\
  \citenamefont {Torner}}]{milianPRL2018}%
  \BibitemOpen
  \bibfield  {author} {\bibinfo {author} {\bibfnamefont {C.}~\bibnamefont
  {Mili{\'a}n}}, \bibinfo {author} {\bibfnamefont {Y.~V.}\ \bibnamefont
  {Kartashov}}, \bibinfo {author} {\bibfnamefont {D.~V.}\ \bibnamefont
  {Skryabin}},\ and\ \bibinfo {author} {\bibfnamefont {L.}~\bibnamefont
  {Torner}},\ }\bibfield  {title} {\bibinfo {title} {Clusters of cavity
  solitons bounded by conical radiation},\ }\href@noop {} {\bibfield  {journal}
  {\bibinfo  {journal} {Phys. Rev. Lett.}\ }\textbf {\bibinfo {volume} {121}},\
  \bibinfo {pages} {103903} (\bibinfo {year} {2018})}\BibitemShut {NoStop}%
\bibitem [{\citenamefont {Ivars}\ \emph {et~al.}(2021)\citenamefont {Ivars},
  \citenamefont {Kartashov}, \citenamefont {Torner}, \citenamefont {Conejero},\
  and\ \citenamefont {Milián}}]{IvarsPRL}%
  \BibitemOpen
  \bibfield  {author} {\bibinfo {author} {\bibfnamefont {S.~B.}\ \bibnamefont
  {Ivars}}, \bibinfo {author} {\bibfnamefont {Y.~V.}\ \bibnamefont
  {Kartashov}}, \bibinfo {author} {\bibfnamefont {L.}~\bibnamefont {Torner}},
  \bibinfo {author} {\bibfnamefont {J.~A.}\ \bibnamefont {Conejero}},\ and\
  \bibinfo {author} {\bibfnamefont {C.}~\bibnamefont {Milián}},\ }\bibfield
  {title} {\bibinfo {title} {Reversible {Self}-{Replication} of
  {Spatiotemporal} {Kerr} {Cavity} {Patterns}},\ }\href
  {https://doi.org/10.1103/PhysRevLett.126.063903} {\bibfield  {journal}
  {\bibinfo  {journal} {Phys. Rev. Lett.}\ }\textbf {\bibinfo {volume} {126}},\
  \bibinfo {pages} {063903} (\bibinfo {year} {2021})}\BibitemShut {NoStop}%
\bibitem [{\citenamefont {Okawachi}\ \emph {et~al.}(2014)\citenamefont
  {Okawachi}, \citenamefont {Lamont}, \citenamefont {Luke}, \citenamefont
  {Carvalho}, \citenamefont {Yu}, \citenamefont {Lipson},\ and\ \citenamefont
  {Gaeta}}]{Okawachi_OL_14}%
  \BibitemOpen
  \bibfield  {author} {\bibinfo {author} {\bibfnamefont {Y.}~\bibnamefont
  {Okawachi}}, \bibinfo {author} {\bibfnamefont {M.~R.~E.}\ \bibnamefont
  {Lamont}}, \bibinfo {author} {\bibfnamefont {K.}~\bibnamefont {Luke}},
  \bibinfo {author} {\bibfnamefont {D.~O.}\ \bibnamefont {Carvalho}}, \bibinfo
  {author} {\bibfnamefont {M.}~\bibnamefont {Yu}}, \bibinfo {author}
  {\bibfnamefont {M.}~\bibnamefont {Lipson}},\ and\ \bibinfo {author}
  {\bibfnamefont {A.~L.}\ \bibnamefont {Gaeta}},\ }\bibfield  {title} {\bibinfo
  {title} {Bandwidth shaping of microresonator-based frequency combs via
  dispersion engineering},\ }\href {https://doi.org/10.1364/OL.39.003535}
  {\bibfield  {journal} {\bibinfo  {journal} {Opt. Lett.}\ }\textbf {\bibinfo
  {volume} {39}},\ \bibinfo {pages} {3535} (\bibinfo {year}
  {2014})}\BibitemShut {NoStop}%
\bibitem [{\citenamefont {Pfeiffer}\ \emph {et~al.}(2017)\citenamefont
  {Pfeiffer}, \citenamefont {Herkommer}, \citenamefont {Liu}, \citenamefont
  {Guo}, \citenamefont {Karpov}, \citenamefont {Lucas}, \citenamefont
  {Zervas},\ and\ \citenamefont {Kippenberg}}]{Pfeiffer_optica_17}%
  \BibitemOpen
  \bibfield  {author} {\bibinfo {author} {\bibfnamefont {M.~H.~P.}\
  \bibnamefont {Pfeiffer}}, \bibinfo {author} {\bibfnamefont {C.}~\bibnamefont
  {Herkommer}}, \bibinfo {author} {\bibfnamefont {J.}~\bibnamefont {Liu}},
  \bibinfo {author} {\bibfnamefont {H.}~\bibnamefont {Guo}}, \bibinfo {author}
  {\bibfnamefont {M.}~\bibnamefont {Karpov}}, \bibinfo {author} {\bibfnamefont
  {E.}~\bibnamefont {Lucas}}, \bibinfo {author} {\bibfnamefont
  {M.}~\bibnamefont {Zervas}},\ and\ \bibinfo {author} {\bibfnamefont {T.~J.}\
  \bibnamefont {Kippenberg}},\ }\bibfield  {title} {\bibinfo {title}
  {Octave-spanning dissipative kerr soliton frequency combs in si3n4
  microresonators},\ }\href {https://doi.org/10.1364/OPTICA.4.000684}
  {\bibfield  {journal} {\bibinfo  {journal} {Optica}\ }\textbf {\bibinfo
  {volume} {4}},\ \bibinfo {pages} {684} (\bibinfo {year} {2017})}\BibitemShut
  {NoStop}%
\bibitem [{\citenamefont {H{\"o}{\"o}k}\ and\ \citenamefont
  {Karlsson}(1993)}]{karlsFOD93}%
  \BibitemOpen
  \bibfield  {author} {\bibinfo {author} {\bibfnamefont {A.}~\bibnamefont
  {H{\"o}{\"o}k}}\ and\ \bibinfo {author} {\bibfnamefont {M.}~\bibnamefont
  {Karlsson}},\ }\bibfield  {title} {\bibinfo {title} {Ultrashort solitons at
  the minimum-dispersion wavelength: effects of fourth-order dispersion},\
  }\href@noop {} {\bibfield  {journal} {\bibinfo  {journal} {Opt. Lett.}\
  }\textbf {\bibinfo {volume} {18}},\ \bibinfo {pages} {1388} (\bibinfo {year}
  {1993})}\BibitemShut {NoStop}%
\bibitem [{\citenamefont {Blanco-Redondo}\ \emph {et~al.}(2016)\citenamefont
  {Blanco-Redondo}, \citenamefont {de~Sterke}, \citenamefont {Sipe},
  \citenamefont {Krauss}, \citenamefont {Eggleton},\ and\ \citenamefont
  {Husko}}]{redNC}%
  \BibitemOpen
  \bibfield  {author} {\bibinfo {author} {\bibfnamefont {A.}~\bibnamefont
  {Blanco-Redondo}}, \bibinfo {author} {\bibfnamefont {C.~M.}\ \bibnamefont
  {de~Sterke}}, \bibinfo {author} {\bibfnamefont {J.~E.}\ \bibnamefont {Sipe}},
  \bibinfo {author} {\bibfnamefont {T.~F.}\ \bibnamefont {Krauss}}, \bibinfo
  {author} {\bibfnamefont {B.~J.}\ \bibnamefont {Eggleton}},\ and\ \bibinfo
  {author} {\bibfnamefont {C.}~\bibnamefont {Husko}},\ }\bibfield  {title}
  {\bibinfo {title} {Pure-quartic solitons},\ }\href
  {https://doi.org/10.1038/ncomms10427} {\bibfield  {journal} {\bibinfo
  {journal} {Nature Communications}\ }\textbf {\bibinfo {volume} {7}},\
  \bibinfo {pages} {10427} (\bibinfo {year} {2016})}\BibitemShut {NoStop}%
\bibitem [{\citenamefont {Runge}\ \emph {et~al.}(2020)\citenamefont {Runge},
  \citenamefont {Hudson}, \citenamefont {Tam}, \citenamefont {de~Sterke},\ and\
  \citenamefont {Blanco-Redondo}}]{redNP}%
  \BibitemOpen
  \bibfield  {author} {\bibinfo {author} {\bibfnamefont {A.~F.~J.}\
  \bibnamefont {Runge}}, \bibinfo {author} {\bibfnamefont {D.~D.}\ \bibnamefont
  {Hudson}}, \bibinfo {author} {\bibfnamefont {K.~K.}\ \bibnamefont {Tam}},
  \bibinfo {author} {\bibfnamefont {C.~M.}\ \bibnamefont {de~Sterke}},\ and\
  \bibinfo {author} {\bibfnamefont {A.}~\bibnamefont {Blanco-Redondo}},\
  }\bibfield  {title} {\bibinfo {title} {The pure-quartic soliton laser},\
  }\href {https://doi.org/10.1038/s41566-020-0629-6} {\bibfield  {journal}
  {\bibinfo  {journal} {Nature Photonics}\ }\textbf {\bibinfo {volume} {14}},\
  \bibinfo {pages} {492} (\bibinfo {year} {2020})}\BibitemShut {NoStop}%
\bibitem [{\citenamefont {Taheri}\ and\ \citenamefont
  {Matsko}(2019)}]{tahOL19}%
  \BibitemOpen
  \bibfield  {author} {\bibinfo {author} {\bibfnamefont {H.}~\bibnamefont
  {Taheri}}\ and\ \bibinfo {author} {\bibfnamefont {A.~B.}\ \bibnamefont
  {Matsko}},\ }\bibfield  {title} {\bibinfo {title} {Quartic dissipative
  solitons in optical {Kerr} cavities},\ }\href
  {https://doi.org/10.1364/OL.44.003086} {\bibfield  {journal} {\bibinfo
  {journal} {Optics Letters}\ }\textbf {\bibinfo {volume} {44}},\ \bibinfo
  {pages} {3086} (\bibinfo {year} {2019})}\BibitemShut {NoStop}%
\bibitem [{\citenamefont {Bao}\ \emph {et~al.}(2017)\citenamefont {Bao},
  \citenamefont {Taheri}, \citenamefont {Zhang}, \citenamefont {Matsko},
  \citenamefont {Yan}, \citenamefont {Liao}, \citenamefont {Maleki},\ and\
  \citenamefont {Willner}}]{Bao_josab_2017}%
  \BibitemOpen
  \bibfield  {author} {\bibinfo {author} {\bibfnamefont {C.}~\bibnamefont
  {Bao}}, \bibinfo {author} {\bibfnamefont {H.}~\bibnamefont {Taheri}},
  \bibinfo {author} {\bibfnamefont {L.}~\bibnamefont {Zhang}}, \bibinfo
  {author} {\bibfnamefont {A.}~\bibnamefont {Matsko}}, \bibinfo {author}
  {\bibfnamefont {Y.}~\bibnamefont {Yan}}, \bibinfo {author} {\bibfnamefont
  {P.}~\bibnamefont {Liao}}, \bibinfo {author} {\bibfnamefont {L.}~\bibnamefont
  {Maleki}},\ and\ \bibinfo {author} {\bibfnamefont {A.~E.}\ \bibnamefont
  {Willner}},\ }\bibfield  {title} {\bibinfo {title} {High-order dispersion in
  kerr comb oscillators},\ }\href@noop {} {\bibfield  {journal} {\bibinfo
  {journal} {J. Opt. Soc. Am. B}\ }\textbf {\bibinfo {volume} {34}},\ \bibinfo
  {pages} {715} (\bibinfo {year} {2017})}\BibitemShut {NoStop}%
\bibitem [{\citenamefont {Liu}\ \emph {et~al.}(2021)\citenamefont {Liu},
  \citenamefont {Yao},\ and\ \citenamefont {Yang}}]{yangOL21}%
  \BibitemOpen
  \bibfield  {author} {\bibinfo {author} {\bibfnamefont {K.}~\bibnamefont
  {Liu}}, \bibinfo {author} {\bibfnamefont {S.}~\bibnamefont {Yao}},\ and\
  \bibinfo {author} {\bibfnamefont {C.}~\bibnamefont {Yang}},\ }\bibfield
  {title} {\bibinfo {title} {Raman pure quartic solitons in {Kerr}
  microresonators},\ }\href {https://doi.org/10.1364/OL.415434} {\bibfield
  {journal} {\bibinfo  {journal} {Optics Letters}\ }\textbf {\bibinfo {volume}
  {46}},\ \bibinfo {pages} {993} (\bibinfo {year} {2021})}\BibitemShut
  {NoStop}%
\bibitem [{\citenamefont {Lugiato}\ and\ \citenamefont
  {Lefever}(1987)}]{lugiato_spatial_1987}%
  \BibitemOpen
  \bibfield  {author} {\bibinfo {author} {\bibfnamefont {L.~A.}\ \bibnamefont
  {Lugiato}}\ and\ \bibinfo {author} {\bibfnamefont {R.}~\bibnamefont
  {Lefever}},\ }\bibfield  {title} {\bibinfo {title} {Spatial {Dissipative}
  {Structures} in {Passive} {Optical} {Systems}},\ }\href
  {https://doi.org/10.1103/PhysRevLett.58.2209} {\bibfield  {journal} {\bibinfo
   {journal} {Physical Review Letters}\ }\textbf {\bibinfo {volume} {58}},\
  \bibinfo {pages} {2209} (\bibinfo {year} {1987})}\BibitemShut {NoStop}%
\bibitem [{\citenamefont {Hooper}\ \emph {et~al.}(2011)\citenamefont {Hooper},
  \citenamefont {Mosley}, \citenamefont {Muir}, \citenamefont {Wadsworth},\
  and\ \citenamefont {Knight}}]{hooper}%
  \BibitemOpen
  \bibfield  {author} {\bibinfo {author} {\bibfnamefont {L.~E.}\ \bibnamefont
  {Hooper}}, \bibinfo {author} {\bibfnamefont {P.~J.}\ \bibnamefont {Mosley}},
  \bibinfo {author} {\bibfnamefont {A.~C.}\ \bibnamefont {Muir}}, \bibinfo
  {author} {\bibfnamefont {W.~J.}\ \bibnamefont {Wadsworth}},\ and\ \bibinfo
  {author} {\bibfnamefont {J.~C.}\ \bibnamefont {Knight}},\ }\bibfield  {title}
  {\bibinfo {title} {Coherent supercontinuum generation in photonic crystal
  fiber with all-normal group velocity dispersion},\ }\href
  {https://doi.org/10.1364/OE.19.004902} {\bibfield  {journal} {\bibinfo
  {journal} {Optics Express}\ }\textbf {\bibinfo {volume} {19}},\ \bibinfo
  {pages} {4902} (\bibinfo {year} {2011})}\BibitemShut {NoStop}%
\bibitem [{\citenamefont {Heidt}(2010)}]{heidt}%
  \BibitemOpen
  \bibfield  {author} {\bibinfo {author} {\bibfnamefont {A.~M.}\ \bibnamefont
  {Heidt}},\ }\bibfield  {title} {\bibinfo {title} {Pulse preserving flat-top
  supercontinuum generation in all-normal dispersion photonic crystal fibers},\
  }\href {https://doi.org/10.1364/JOSAB.27.000550} {\bibfield  {journal}
  {\bibinfo  {journal} {J. Opt. Soc. Am. B}\ }\textbf {\bibinfo {volume}
  {27}},\ \bibinfo {pages} {550} (\bibinfo {year} {2010})}\BibitemShut
  {NoStop}%
\bibitem [{\citenamefont {Yao}\ \emph {et~al.}(2021)\citenamefont {Yao},
  \citenamefont {Liu},\ and\ \citenamefont {Yang}}]{yaoOE21}%
  \BibitemOpen
  \bibfield  {author} {\bibinfo {author} {\bibfnamefont {S.}~\bibnamefont
  {Yao}}, \bibinfo {author} {\bibfnamefont {K.}~\bibnamefont {Liu}},\ and\
  \bibinfo {author} {\bibfnamefont {C.}~\bibnamefont {Yang}},\ }\bibfield
  {title} {\bibinfo {title} {Pure quartic solitons in dispersion-engineered
  aluminum nitride micro-cavities},\ }\href {https://doi.org/10.1364/OE.418538}
  {\bibfield  {journal} {\bibinfo  {journal} {Optics Express}\ }\textbf
  {\bibinfo {volume} {29}},\ \bibinfo {pages} {8312} (\bibinfo {year}
  {2021})}\BibitemShut {NoStop}%
\bibitem [{\citenamefont {Woods}\ and\ \citenamefont
  {Champneys}(1999)}]{woods_heteroclinic_1999}%
  \BibitemOpen
  \bibfield  {author} {\bibinfo {author} {\bibfnamefont {P.~D.}\ \bibnamefont
  {Woods}}\ and\ \bibinfo {author} {\bibfnamefont {A.~R.}\ \bibnamefont
  {Champneys}},\ }\bibfield  {title} {\bibinfo {title} {Heteroclinic tangles
  and homoclinic snaking in the unfolding of a degenerate reversible
  {Hamiltonian}–{Hopf} bifurcation},\ }\href
  {https://doi.org/10.1016/S0167-2789(98)00309-1} {\bibfield  {journal}
  {\bibinfo  {journal} {Physica D: Nonlinear Phenomena}\ }\textbf {\bibinfo
  {volume} {129}},\ \bibinfo {pages} {147} (\bibinfo {year}
  {1999})}\BibitemShut {NoStop}%
\bibitem [{\citenamefont {Champneys}(1998)}]{champneys_homoclinic_1998}%
  \BibitemOpen
  \bibfield  {author} {\bibinfo {author} {\bibfnamefont {A.~R.}\ \bibnamefont
  {Champneys}},\ }\bibfield  {title} {\bibinfo {title} {Homoclinic orbits in
  reversible systems and their applications in mechanics, fluids and optics},\
  }\href {https://doi.org/10.1016/S0167-2789(97)00209-1} {\bibfield  {journal}
  {\bibinfo  {journal} {Physica D: Nonlinear Phenomena}\ }\bibinfo {series}
  {Proceedings of the {Workshop} on {Time}-{Reversal} {Symmetry} in {Dynamical}
  {Systems}},\ \textbf {\bibinfo {volume} {112}},\ \bibinfo {pages} {158}
  (\bibinfo {year} {1998})}\BibitemShut {NoStop}%
\bibitem [{\citenamefont {Parra-Rivas}\ \emph
  {et~al.}(2018{\natexlab{a}})\citenamefont {Parra-Rivas}, \citenamefont
  {Gomila}, \citenamefont {Gelens},\ and\ \citenamefont
  {Knobloch}}]{parra-rivas_bifurcation_2018}%
  \BibitemOpen
  \bibfield  {author} {\bibinfo {author} {\bibfnamefont {P.}~\bibnamefont
  {Parra-Rivas}}, \bibinfo {author} {\bibfnamefont {D.}~\bibnamefont {Gomila}},
  \bibinfo {author} {\bibfnamefont {L.}~\bibnamefont {Gelens}},\ and\ \bibinfo
  {author} {\bibfnamefont {E.}~\bibnamefont {Knobloch}},\ }\bibfield  {title}
  {\bibinfo {title} {Bifurcation structure of localized states in the
  {Lugiato}-{Lefever} equation with anomalous dispersion},\ }\href
  {https://doi.org/10.1103/PhysRevE.97.042204} {\bibfield  {journal} {\bibinfo
  {journal} {Physical Review E}\ }\textbf {\bibinfo {volume} {97}},\ \bibinfo
  {pages} {042204} (\bibinfo {year} {2018}{\natexlab{a}})}\BibitemShut
  {NoStop}%
\bibitem [{\citenamefont {Qi}\ \emph {et~al.}(2019)\citenamefont {Qi},
  \citenamefont {Wang}, \citenamefont {Jaramillo-Villegas}, \citenamefont {Qi},
  \citenamefont {Weiner}, \citenamefont {D’Aguanno}, \citenamefont
  {Carruthers},\ and\ \citenamefont {Menyuk}}]{qi_dissipative_2019}%
  \BibitemOpen
  \bibfield  {author} {\bibinfo {author} {\bibfnamefont {Z.}~\bibnamefont
  {Qi}}, \bibinfo {author} {\bibfnamefont {S.}~\bibnamefont {Wang}}, \bibinfo
  {author} {\bibfnamefont {J.}~\bibnamefont {Jaramillo-Villegas}}, \bibinfo
  {author} {\bibfnamefont {M.}~\bibnamefont {Qi}}, \bibinfo {author}
  {\bibfnamefont {A.~M.}\ \bibnamefont {Weiner}}, \bibinfo {author}
  {\bibfnamefont {G.}~\bibnamefont {D’Aguanno}}, \bibinfo {author}
  {\bibfnamefont {T.~F.}\ \bibnamefont {Carruthers}},\ and\ \bibinfo {author}
  {\bibfnamefont {C.~R.}\ \bibnamefont {Menyuk}},\ }\bibfield  {title}
  {\bibinfo {title} {Dissipative cnoidal waves ({Turing} rolls) and the soliton
  limit in microring resonators},\ }\href
  {https://doi.org/10.1364/OPTICA.6.001220} {\bibfield  {journal} {\bibinfo
  {journal} {Optica}\ }\textbf {\bibinfo {volume} {6}},\ \bibinfo {pages}
  {1220} (\bibinfo {year} {2019})}\BibitemShut {NoStop}%
\bibitem [{\citenamefont {Knobloch}\ and\ \citenamefont
  {Wagenknecht}(2005)}]{knobloch_homoclinic_2005}%
  \BibitemOpen
  \bibfield  {author} {\bibinfo {author} {\bibfnamefont {J.}~\bibnamefont
  {Knobloch}}\ and\ \bibinfo {author} {\bibfnamefont {T.}~\bibnamefont
  {Wagenknecht}},\ }\bibfield  {title} {\bibinfo {title} {Homoclinic snaking
  near a heteroclinic cycle in reversible systems},\ }\href
  {https://doi.org/10.1016/j.physd.2005.04.018} {\bibfield  {journal} {\bibinfo
   {journal} {Physica D: Nonlinear Phenomena}\ }\textbf {\bibinfo {volume}
  {206}},\ \bibinfo {pages} {82} (\bibinfo {year} {2005})}\BibitemShut
  {NoStop}%
\bibitem [{\citenamefont {Lobanov}\ \emph {et~al.}(2015)\citenamefont
  {Lobanov}, \citenamefont {Lihachev}, \citenamefont {Kippenberg},\ and\
  \citenamefont {Gorodetsky}}]{lobanov_frequency_2015}%
  \BibitemOpen
  \bibfield  {author} {\bibinfo {author} {\bibfnamefont {V.~E.}\ \bibnamefont
  {Lobanov}}, \bibinfo {author} {\bibfnamefont {G.}~\bibnamefont {Lihachev}},
  \bibinfo {author} {\bibfnamefont {T.~J.}\ \bibnamefont {Kippenberg}},\ and\
  \bibinfo {author} {\bibfnamefont {M.~L.}\ \bibnamefont {Gorodetsky}},\
  }\bibfield  {title} {\bibinfo {title} {Frequency combs and platicons in
  optical microresonators with normal {GVD}},\ }\href
  {https://doi.org/10.1364/OE.23.007713} {\bibfield  {journal} {\bibinfo
  {journal} {Optics Express}\ }\textbf {\bibinfo {volume} {23}},\ \bibinfo
  {pages} {7713} (\bibinfo {year} {2015})}\BibitemShut {NoStop}%
\bibitem [{\citenamefont {Parra-Rivas}\ \emph {et~al.}(2016)\citenamefont
  {Parra-Rivas}, \citenamefont {Knobloch}, \citenamefont {Gomila},\ and\
  \citenamefont {Gelens}}]{parra-rivas_dark_2016}%
  \BibitemOpen
  \bibfield  {author} {\bibinfo {author} {\bibfnamefont {P.}~\bibnamefont
  {Parra-Rivas}}, \bibinfo {author} {\bibfnamefont {E.}~\bibnamefont
  {Knobloch}}, \bibinfo {author} {\bibfnamefont {D.}~\bibnamefont {Gomila}},\
  and\ \bibinfo {author} {\bibfnamefont {L.}~\bibnamefont {Gelens}},\
  }\bibfield  {title} {\bibinfo {title} {Dark solitons in the
  {Lugiato}-{Lefever} equation with normal dispersion},\ }\href
  {https://doi.org/10.1103/PhysRevA.93.063839} {\bibfield  {journal} {\bibinfo
  {journal} {Physical Review A}\ }\textbf {\bibinfo {volume} {93}},\ \bibinfo
  {pages} {063839} (\bibinfo {year} {2016})}\BibitemShut {NoStop}%
\bibitem [{\citenamefont {Parra-Rivas}\ \emph {et~al.}(2017)\citenamefont
  {Parra-Rivas}, \citenamefont {Gomila},\ and\ \citenamefont
  {Gelens}}]{parra-rivas_coexistence_2017}%
  \BibitemOpen
  \bibfield  {author} {\bibinfo {author} {\bibfnamefont {P.}~\bibnamefont
  {Parra-Rivas}}, \bibinfo {author} {\bibfnamefont {D.}~\bibnamefont
  {Gomila}},\ and\ \bibinfo {author} {\bibfnamefont {L.}~\bibnamefont
  {Gelens}},\ }\bibfield  {title} {\bibinfo {title} {Coexistence of stable
  dark- and bright-soliton {Kerr} combs in normal-dispersion resonators},\
  }\href {https://doi.org/10.1103/PhysRevA.95.053863} {\bibfield  {journal}
  {\bibinfo  {journal} {Physical Review A}\ }\textbf {\bibinfo {volume} {95}},\
  \bibinfo {pages} {053863} (\bibinfo {year} {2017})}\BibitemShut {NoStop}%
\bibitem [{\citenamefont {Li}\ \emph {et~al.}(2020)\citenamefont {Li},
  \citenamefont {Xu}, \citenamefont {Coen}, \citenamefont {Murdoch},\ and\
  \citenamefont {Erkintalo}}]{li_experimental_2020}%
  \BibitemOpen
  \bibfield  {author} {\bibinfo {author} {\bibfnamefont {Z.}~\bibnamefont
  {Li}}, \bibinfo {author} {\bibfnamefont {Y.}~\bibnamefont {Xu}}, \bibinfo
  {author} {\bibfnamefont {S.}~\bibnamefont {Coen}}, \bibinfo {author}
  {\bibfnamefont {S.~G.}\ \bibnamefont {Murdoch}},\ and\ \bibinfo {author}
  {\bibfnamefont {M.}~\bibnamefont {Erkintalo}},\ }\bibfield  {title} {\bibinfo
  {title} {Experimental observations of bright dissipative cavity solitons and
  their collapsed snaking in a kerr resonator with normal dispersion driving},\
  }\href@noop {} {\bibfield  {journal} {\bibinfo  {journal} {Optica}\ }\textbf
  {\bibinfo {volume} {7}},\ \bibinfo {pages} {1195} (\bibinfo {year}
  {2020})}\BibitemShut {NoStop}%
\bibitem [{\citenamefont {Talla~Mbé}\ \emph {et~al.}(2017)\citenamefont
  {Talla~Mbé}, \citenamefont {Milián},\ and\ \citenamefont {Chembo}}]{mbe17}%
  \BibitemOpen
  \bibfield  {author} {\bibinfo {author} {\bibfnamefont {J.~H.}\ \bibnamefont
  {Talla~Mbé}}, \bibinfo {author} {\bibfnamefont {C.}~\bibnamefont
  {Milián}},\ and\ \bibinfo {author} {\bibfnamefont {Y.~K.}\ \bibnamefont
  {Chembo}},\ }\bibfield  {title} {\bibinfo {title} {Existence and switching
  behavior of bright and dark {Kerr} solitons in whispering-gallery mode
  resonators with zero group-velocity dispersion},\ }\href
  {https://doi.org/10.1140/epjd/e2017-80132-8} {\bibfield  {journal} {\bibinfo
  {journal} {The European Physical Journal D}\ }\textbf {\bibinfo {volume}
  {71}},\ \bibinfo {pages} {196} (\bibinfo {year} {2017})}\BibitemShut
  {NoStop}%
\bibitem [{\citenamefont {Wiggins}(2003)}]{wiggins_introduction_2003}%
  \BibitemOpen
  \bibfield  {author} {\bibinfo {author} {\bibfnamefont {S.}~\bibnamefont
  {Wiggins}},\ }\href {https://www.springer.com/gp/book/9780387001777} {\emph
  {\bibinfo {title} {Introduction to {Applied} {Nonlinear} {Dynamical}
  {Systems} and {Chaos}}}},\ \bibinfo {edition} {2nd}\ ed.,\ Texts in {Applied}
  {Mathematics}\ (\bibinfo  {publisher} {Springer-Verlag},\ \bibinfo {address}
  {New York},\ \bibinfo {year} {2003})\BibitemShut {NoStop}%
\bibitem [{\citenamefont {Parra-Rivas}\ \emph
  {et~al.}(2018{\natexlab{b}})\citenamefont {Parra-Rivas}, \citenamefont
  {Gomila}, \citenamefont {Gelens},\ and\ \citenamefont
  {Knobloch}}]{parra-rivas_bifurcation_2018-1}%
  \BibitemOpen
  \bibfield  {author} {\bibinfo {author} {\bibfnamefont {P.}~\bibnamefont
  {Parra-Rivas}}, \bibinfo {author} {\bibfnamefont {D.}~\bibnamefont {Gomila}},
  \bibinfo {author} {\bibfnamefont {L.}~\bibnamefont {Gelens}},\ and\ \bibinfo
  {author} {\bibfnamefont {E.}~\bibnamefont {Knobloch}},\ }\bibfield  {title}
  {\bibinfo {title} {Bifurcation structure of periodic patterns in the
  {Lugiato}-{Lefever} equation with anomalous dispersion},\ }\href
  {https://doi.org/10.1103/PhysRevE.98.042212} {\bibfield  {journal} {\bibinfo
  {journal} {Physical Review E}\ }\textbf {\bibinfo {volume} {98}},\ \bibinfo
  {pages} {042212} (\bibinfo {year} {2018}{\natexlab{b}})}\BibitemShut
  {NoStop}%
\end{thebibliography}%


%

\end{document}